\documentclass[showpacs]{revtex4}
\usepackage{amsmath}
\usepackage{amssymb}
\usepackage{amsfonts}
\usepackage{graphicx}
\usepackage{color}
\usepackage{dcolumn}

\begin{document}
\newcolumntype{.}{D{.}{.}{-1}}

\title{
Anisotropic neutron stars in $R^2$ gravity
}
\author{Vladimir Folomeev}
\email{vfolomeev@mail.ru}
\affiliation{
	Institute of Experimental and Theoretical Physics,  Al-Farabi Kazakh National University, Almaty 050040, Kazakhstan;
}
\affiliation{
	National Nanotechnology Laboratory of Open Type,  Al-Farabi Kazakh National University, Almaty 050040, Kazakhstan
}
\affiliation{
	Institute of Physicotechnical Problems and Material Science of the 
National Academy of Sciences of the Kyrgyz Republic, 265 a, Chui Street, Bishkek 720071,  Kyrgyzstan
}

\begin{abstract}
We consider static neutron stars within the framework of $R^2$ gravity.
The neutron fluid is described by three different types of realistic equations of state (soft, moderately stiff, and stiff).
Using the observational data on the neutron star mass-radius relation,
 it is demonstrated that the characteristics of
 the objects supported by the isotropic fluid agree with the observations only if one uses the soft
equation of state. We show that the inclusion of the fluid anisotropy
enables one also to employ 
more stiff equations of state to model configurations that will satisfy the observational constraints sufficiently.
Also, using the standard thin accretion disk model, we demonstrate potentially observable differences, which allow us to distinguish the neutron stars
constructed within the modified gravity framework from those described in Einstein's general relativity.
\end{abstract}


\maketitle

\vspace{-.5cm}

\section{Introduction}
Neutron stars (NSs) are good objects for testing different theoretical models of matter under extreme physical conditions.
In fact, superhigh densities (of the order of nuclear density) and pressures are typical for internal regions of NSs.
Such matter cannot be created in a laboratory;  its properties and detailed composition are not completely known at present.
For its description, one can only employ theoretical models.
The verification of such models is performed by analyzing and interpreting the results of astronomical observations
with subsequent refinement of original theoretical models~\cite{Potekhin:2011xe}.

On the other hand, the physical characteristics of NSs are also largely determined by their own strong gravitational fields.
A description of the latter can be performed within the framework of various theories of gravity.
Usually, a consideration of NSs is carried out in Einstein's general relativity (GR), within which
significant progress has already been made in constructing theoretical
models that adequately represent the observational properties  of NSs
(see, e.g., Ref.~\cite{HPY}).

However, GR is not the only possible theory of gravity. After the discovery of the accelerated expansion of the present Universe,
various modified gravity theories (MGTs) extending GR
have found many applications in describing the current Universe.
One of the main advantages of such theories is that,
in contrast to GR, they do not require the introduction of  any special exotic forms of matter (dark energy). 
In the simplest case, the modification of GR reduces to the replacement of the Einstein gravitational Lagrangian $\sim R$ by the
modified Lagrangian $\sim f(R)$, where $f(R)$ is some function of the scalar curvature $R$.
Such  MGTs had initially been applied for the description of
the evolution of the very early Universe, but it was shown in recent years that they can also be successfully applied to model  various
cosmological aspects of the present Universe (for a general review on the subject, see, e.g., Refs.~\cite{Nojiri:2010wj,DeFelice:2010aj,Nojiri:2017ncd}).

When one considers smaller (astrophysical) scales typical for stars, the effects of modification of gravity can also play a significant role.
In particular, within the framework of $f(R)$ gravity,
one can construct relativistic stars~\cite{rel_star_f_R_1,rel_star_f_R_2}
or such exotic objects as wormholes~\cite{wh_f_R_1,wh_f_R_2}. However, the effects of MGTs may also manifest themselves in considering less
exotic objects like neutron stars~\cite{Cooney:2009rr,Arapoglu:2010rz,Orellana:2013gn,Alavirad:2013paa,Astashenok:2013vza,Ganguly:2013taa}.
Modification of gravity can affect a number of important physical characteristics of NSs which can, in principle, be directly verified observationally.
Among them are the mass-radius ($M-{\cal R}$) relation~\cite{Astashenok:2014pua,Capozziello:2015yza,Astashenok:2017dpo},
the properties of electromagnetic radiation from the surface of accretion disks~\cite{Staykov:2016dzn},
and the structure of internal and external magnetic fields~\cite{Cheoun:2013tsa,Astashenok:2014gda,Astashenok:2014nua,Bakirova:2016ffk}.
Considering such objects within the framework of different types of
$f(R)$ gravities and using  various equations of state (EoSs) for neutron matter, one can
reveal the allowed forms of $f(R)$ and EoSs satisfying the observational constraints.

Regardless of the theory of gravity used to model NSs, their properties and structure are strictly correlated with an EoS
of matter supporting the stars. The literature in the field offers dozens of different EoSs
that are assumed to be suitable for modeling NSs~\cite{HPY}.
It is evident that the choice of the most realistic EoSs from that set should be carried out, in particular, on the basis of results of astronomical observations.
This applies, for instance, to measurements of masses of NSs in binary systems.
Such measurements ensure the greatest accuracy and yield a mass range from
 $\sim 1.35 M_\odot$ (for the binary radio pulsars of Ref.~\cite{Thorsett:1998uc}) to $\sim 2 M_\odot$ (radio pulsars J~$0751 + 1807$~\cite{Nice:2005fi}
and PSR~J~$1614 - 2230$~\cite{Demorest:2010bx}).

On one hand, the aforementioned observational constraints on the masses of NSs enable one to exclude some of the EoSs.
In particular, this applies to
stiff EoSs, which usually give the $M-{\cal R}$ relations which do not satisfy the observational constraints (see in Sec.~\ref{num_calc} below).
However, it should be emphasized that investigations of the structure of NSs
are usually carried out under the supposition that their matter is described by an isotropic perfect-fluid EoS,
i.e., by a fluid obeying Pascal's law when the radial and tangential components of the pressure are equal to each other.
However, due to the presence of ultrastrong magnetic fields and extremely large densities  and pressures in the internal regions of this type of star,
such a description cannot be always considered completely satisfactory.
In particular, one can expect the appearance of unequal principal stresses in the
neutron fluid caused by the presence of strong magnetic fields
(see Refs.~\cite{Chaichian:1999gd,PerezMartinez:2007kw,Ferrer:2010wz} and references therein).
Among the other possible reasons for the appearance of the
anisotropy in superdense matter might be
nuclear interactions~\cite{Rud1972},  pion condensation~\cite{Saw1972},
 some kinds of phase transitions~\cite{Sok1980}, and viscosity effects~\cite{Ivanov:2010my}.
Regardless of the specific nature of the fluid anisotropy,
its presence may lead to significant changes in the characteristics of relativistic stars, as demonstrated, for
instance, in Refs.~\cite{BL1974,HH1975_1,HH1975_2,anis_stars_1,anis_stars_2,anis_stars_3,anis_stars_4,Horvat:2010xf,Herrera:2013fja}.
In particular, the presence of the anisotropy enables one to increase or decrease the mass of configurations constructed with different EoSs.
This allows the possibility of obtaining objects satisfying the observational constraints.

According to the literature mentioned above, the studies of the anisotropic systems are usually carried out in Einstein's gravity.
Within the framework of extended theories of gravity,
anisotropic stars have particularly been considered in Ref.~\cite{Silva:2014fca}, where static and
 slowly rotating objects in the scalar-tensor theory of gravity have been investigated.
To the best of our knowledge, anisotropic stars have still not been studied in $f(R)$ gravity.
The aim of the present work is to
fill this gap. To do so, we will consider the case of the simplest $R^2$ gravity, which is often discussed in the literature as a
viable alternative cosmological model describing the accelerated expansion of the early and present Universe~\cite{DeFelice:2010aj,Nojiri:2010wj,Nojiri:2017ncd}.
Within this theory, the neutron-star's matter will be modeled by three different types of EoSs (soft, moderately stiff, and stiff).

In turn, in modeling the anisotropy of superdense matter, one might expect that it must be determined
by relationships between the components (radial and tangential) of the pressure and the energy density of the fluid. Unfortunately,
at the present time it seems impossible to find a specific form of such relationships from the first-principles theory.
In this connection, the literature in the field offers
 several more or less physically motivated functional relations for the anisotropy, which allow a smooth
transition between isotropic and anisotropic states (for a detailed discussion, see, e.g., Refs. ~\cite{BL1974,Horvat:2010xf}). In the present paper
we will employ two phenomenological models of the anisotropy known from the literature.
Our goal will be to examine the possibility of using the anisotropy to obtain configurations constructed with various EoSs and to satisfy
the current observational data on the $M-{\cal R}$ relation.

We will also consider one more important observational manifestation of NSs associated with a
process of accretion of surrounding matter onto a star. Namely,
 we will study a steady-state accretion process for a geometrically thin and optically thick accretion disc
orbiting NSs. The energy released in such a process may be converted into observable radiation.
Our purpose will be to reveal the differences in the  emitted radiation pattern of isotropic and anisotropic configurations with the same masses
described in GR and in the MGT.

The paper is organized as follows: In Sec.~\ref{statem_prob},
 we describe the problem and derive the corresponding
general equations within the framework of $f(R)$ gravity for the configurations under consideration.
These equations are solved numerically in Sec.~\ref{num_calc} in the particular case of $R^2$ gravity and
when the neutron fluid is described by realistic EoSs.
Comparing the results from GR and the MGT,
we demonstrate the effects
of modified gravity and fluid anisotropy on the  $M-{\cal R}$ relation and on the internal structure of the neutron stars.
Next, to reveal additional observational differences,
in Sec.~\ref{sec_thin_acc_disk}, we consider  the process of thin-disk accretion onto such objects
and compare the energy fluxes emitted  from the disk's surface.
Finally, in Sec.~\ref{concl}, we summarize the obtained results.

\section{Statement of the problem and general equations}
\label{statem_prob}

We consider modified gravity
with the action [the metric signature is $(+,-,-,-)$]
\begin{equation}
\label{action_mod}
S=-\frac{c^4}{16\pi G}\int d^4 x \sqrt{-g} f(R) +S_m,
\end{equation}
where $G$ is the Newtonian gravitational constant,
$f(R)$ is an arbitrary nonlinear function of $R$, and $S_m$ denotes the action
of matter. Note that in the present paper we work strictly in
the Jordan frame, where the matter is minimally coupled to geometry.

The literature in the field offers two approaches to considering NSs within the framework of $f(R)$ gravity: perturbative and nonperturbative.
Within the first approach,
  the deviations from GR are assumed to be small (see, e.g., Ref.~\cite{Cooney:2009rr}).
Then the resulting field equations are  second-order differential equations
with respect to metric functions. Here, we will use a fully self-consistent nonperturbative approach where one seeks
 solutions of exact higher-order differential equations. In this case one can expect that the non-GR
gravitational effects will be dominant; this may result in new consequences, which are absent within the framework of the perturbative approach.

For our purposes, we represent the function $f(R)$  in the form $f(R)=R+\alpha h(R)$, where $h(R)$ is new arbitrary function of $R$
and $\alpha$ is an arbitrary constant. When $\alpha=0$, one recovers Einstein's general relativity.
The corresponding field equations can be obtained by
varying action \eqref{action_mod} with respect to the metric, yielding
\begin{equation}
\label{mod_Ein_eqs_gen}
\left(1+\alpha h_R\right) G_i^k-\frac{1}{2}\alpha\left(h-R\,h_R \right)\delta_i^k+
\alpha \left(\delta_i^k g^{m n}-\delta_i^m g^{k n}\right)\left(h_R\right)_{;m;n}=\frac{8\pi G}{c^4}T_i^k.
\end{equation}
Here $G_i^k\equiv R_i^k-\frac{1}{2}\delta_i^k R$ is the Einstein tensor, $h_R\equiv dh/dR$, and
the semicolon denotes the covariant derivative.

To derive the modified Einstein equations and the equation for the fluid, we choose
the spherically symmetric line element in the form
\begin{equation}
\label{metric_schw}
ds^2=e^{\nu}(dx^0)^2-e^{\lambda}dr^2-r^2 \left(d\Theta^2+\sin^2\Theta\, d\phi^2\right),
\end{equation}
where $\nu$ and $\lambda$ are functions of the radial coordinate $r$ only,
and $x^0=c\, t$ is the time coordinate.

As a matter source in the field equations, we take an anisotropic
fluid, i.e., the fluid for which
 the radial,~$p_r$, and tangential, $p_t$, components of the pressure are not equal to each other.
For such a fluid,
the energy-momentum tensor can be written in the form (see, e.g., Ref.~\cite{Herrera:2013fja})
\begin{equation}
\label{fluid_emt_anis}
T_{i}^k=\left(\varepsilon +p_t\right)u^k u_i-\delta_i^k p_t+\left(p_r-p_t\right)s^k s_i,
\end{equation}
where $\varepsilon$ is the fluid energy density. The radial unit vector  $s^k$ is defined as
$s^k=\left(0, e^{-\lambda/2},0,0\right)$, with $s^k s_k=-1$ and $s^k u_k=0$. Then the energy-momentum tensor contains only the following
nonzero diagonal components: $T_{i }^k=\left(\varepsilon,-p_r, -p_t, -p_t\right)$.

The trace of Eq.~\eqref{mod_Ein_eqs_gen} gives the equation for the scalar curvature
\begin{equation}
\label{scal_cur_eq}
R^{\prime\prime}=-\left[\frac{2}{r}+\frac{1}{2}\left(\nu^\prime-\lambda^\prime\right)\right] R^\prime-\frac{h_{3R}}{h_{2R}}R^{\prime 2}+
\frac{e^\lambda}{3h_{2R}}\left[R\, h_R-2 h-\frac{1}{\alpha}\left(\frac{8\pi G}{c^4}T+R\right)\right],
\end{equation}
where $T$ is the trace of the energy-momentum tensor \eqref{fluid_emt_anis} and the prime denotes differentiation with respect to $r$.

In turn, the
$(^t_t)$ and $(^r_r)$ components of Eq.~\eqref{mod_Ein_eqs_gen} are
\begin{eqnarray}
\label{mod_Einstein-00_gen}
&&\left(1+\alpha h_R\right)
\left[-e^{-\lambda}\left(\frac{1}{r^2}-\frac{\lambda^\prime}{r}\right)+\frac{1}{r^2}\right]-
\alpha\left\{\frac{1}{2}\left(h-h_R R\right)+e^{-\lambda}\left[h_R^{\prime\prime}-\left(\frac{1}{2}\lambda^\prime-\frac{2}{r}\right)h_R^{\prime}\right]
\right\}
=\frac{8\pi G}{c^4} \varepsilon,
 \\
\label{mod_Einstein-11_gen}
&&\left(1+\alpha h_R\right)
\left[-e^{-\lambda}\left(\frac{1}{r^2}+\frac{\nu^\prime}{r}\right)+\frac{1}{r^2}\right]-
\alpha\left[\frac{1}{2}\left(h-h_R R\right)+e^{-\lambda}\left(\frac{1}{2}\nu^\prime+\frac{2}{r}\right)h_R^{\prime}
\right]
=-\frac{8\pi G}{c^4} p_r,
\end{eqnarray}
where  the right-hand sides have been taken from \eqref{fluid_emt_anis}.

Introducing a new function $M(r)$, defined as
\begin{equation}
\label{metr_g11}
e^{-\lambda}=1-\frac{2 G M(r)}{c^2 r},
\end{equation}
Eq.~\eqref{mod_Einstein-00_gen} can be rewritten in the form
\begin{equation}
\label{mass_eq}
\frac{d M}{d r}=\frac{4\pi}{c^2} r^2 \varepsilon-\alpha \frac{c^2}{2 G}r^2
\left\{\frac{8\pi G}{c^4}h_R \varepsilon -\frac{1}{2}\left(h-h_R R\right)-e^{-\lambda}\left[h_R^{\prime\prime}-\left(\frac{1}{2}\lambda^\prime-\frac{2}{r}\right)h_R^{\prime}\right]
\right\}.
\end{equation}

Notice here that when one considers compact configurations in GR (when $\alpha=0$),
the function $M(r)$ plays the role of the current mass enclosed by a sphere
with circumferential radius $r$. Then outside the star
 (where $\varepsilon=0$), $M=\text{const.}$ is the total gravitational mass of the configuration.
  A different situation takes place in the MGT (when $\alpha\neq0$): outside the neutron fluid
the scalar curvature $R\neq 0$. (In the terminology of Ref.~\cite{Astashenok:2017dpo}, the star is surrounded by the gravitational sphere.)
This sphere gives an extra contribution to the gravitational mass measured by a distant observer.
As pointed out in Ref.~\cite{Astashenok:2017dpo}, depending on the sign of $\alpha$, one may find either an asymptotically damped behavior of the metric function
$\lambda$ [and correspondingly of the scalar curvature $R$ and of
the mass function $M(r)$] or its oscillation. In the latter case $M(r)$ from \eqref{metr_g11} cannot already be interpreted as the mass function
that forces one to use another way to define the mass (see Ref.~\cite{Astashenok:2017dpo}). In the present paper we deal only with $\alpha$'s
that ensure the asymptotically damped  behavior of $M(r)$ without oscillations. This enables one to interpret $M(r\to \infty)$ as the total
gravitational mass (see below in Sec.~\ref{num_calc}).

Finally, the $i=r$ component
of the conservation law,
$T^k_{i; k}=0$, yields the  equation
\begin{equation}
\label{conserv_gen}
\frac{d p_r}{d r}=-\frac{1}{2}\left(\varepsilon+p_r\right)\frac{d \nu}{d r}+\frac{2}{r}\left(p_t-p_r\right).
\end{equation}

For a complete description of the  configuration under consideration, the above equations
have to be supplemented by an equation of state for the fluid.
Here, we consider only
a simple barotropic EoS where the pressure is a function of the mass density $\rho$.
In this case, one has two possibilities to specify the EoS. First, it is possible to assign two different EoSs
for the radial and the tangential  components of the pressure, $p_r=p_r(\rho)$
and $p_t=p_t(\rho)$. Second, one can take only one EoS,
say, $p_r=p_r(\rho)$, but, in addition to this, it is possible to assign the function
$\Delta\equiv p_t-p_r$, which appears in Eq.~\eqref{conserv_gen}.
This function is called the anisotropy factor~\cite{Herrera:1985}.

We here employ the second possibility, for which we take two different functions $\Delta$ used in the literature in modeling anisotropic matter at high densities
in strong gravitational fields:
\begin{enumerate}
\itemsep=-0.2pt
\item[(1)] Quasi-local EoS suggested by Horvat {\it et al.} in Ref.~\cite{Horvat:2010xf}:
\begin{equation}
\label{anis_fact_1}
\Delta\equiv p_t-p_r=\lambda_{\text{H}} p_r \mu,
\end{equation}
where $\lambda_{\text{H}}$ is a free parameter that controls the degree of
anisotropy and the function $$\mu=\frac{2 G M(r)}{c^2 r}$$ is called the compactness.

The choice \eqref{anis_fact_1} has the following two particularly
attractive features~\cite{Horvat:2010xf}. First, since as $r\to 0$ the compactness $\mu \sim r^2$,
 the anisotropy factor vanishes at the center (i.e., the fluid becomes isotropic),
and this ensures the regularity of the right-hand side
of Eq.~\eqref{conserv_gen} (other possibilities of obtaining regular solutions without imposing the requirement for the anisotropy to vanish
at the center can be found in Refs.~\cite{HH1975_1,HH1975_2}). Second,
the anisotropy factor given in the form \eqref{anis_fact_1} is important only for
essentially relativistic configurations, for which
$\mu\sim {\cal O}(1)$. This is in accord with the conventional assumption, according to which the fluid anisotropy may manifest itself
only at high densities of matter~\cite{BL1974,HH1975_1,HH1975_2,anis_stars_1,anis_stars_2,anis_stars_3,anis_stars_4,Horvat:2010xf}.

The magnitude of the  anisotropy parameter $\lambda_{\text{H}}$ can be of the order of unity~\cite{Sawyer:1973fv,Nelmes:2012uf},
and the literature in the field offers the range $-2\leq\lambda_{\text{H}}\leq 2$~\cite{Doneva:2012rd,Silva:2014fca,Folomeev:2015aua}.

 \item[(2)] Another form of the anisotropy factor,
 \begin{equation}
\label{anis_fact_2}
 \Delta=\lambda_{\text{BL}} \frac{G}{c^4}\left(\varepsilon+p_r\right)\left(\varepsilon+3 p_r\right)e^\lambda r^2,
 \end{equation}
 has been employed by Bowers and Liang~\cite{BL1974} to describe
  incompressible stars with a constant density. As in the case of the anisotropy factor  from  \eqref{anis_fact_1}, this $\Delta$
 is (in part) gravitationally induced (through the factor $e^\lambda$), but depends nonlinearly on  $p_r$ and $\varepsilon$.
  The  anisotropy parameter $\lambda_{\text{BL}}$ entering here is also of the order of  1 (see, e.g., Ref.~\cite{Silva:2014fca}, where  $-2/3\leq\lambda_{\text{BL}}\leq 2/3$).
\end{enumerate}

\section{Numerical results}
\label{num_calc}

In this section we numerically integrate the  equations of Sec.~\ref{statem_prob}.
To do so, one needs to choose an EoS for the neutron matter. This can be any EoS  used in the literature
to describe matter at high densities and pressures (see, e.g., Refs.~\cite{Potekhin:2011xe,HPY}). We use here
three well-known EoSs: the~soft
FPS EoS, the moderately stiff SLy EoS,  and the stiff BSk21 EoS.
They can be represented  by the corresponding analytical approximations. For example, for the SLy EoS one has
\begin{eqnarray}
\label{EOS_analyt}
\zeta=\frac{a_{1}+a_{2}\xi+a_{3}\xi^3}{1+a_{4}\xi}f(a_{5}(\xi-a_{6}))+(a_{7}+a_{8}\xi)f(a_{9}(a_{10}-\xi)) \nonumber\\
+(a_{11}+a_{12}\xi)f(a_{13}(a_{14}-\xi))+(a_{15}+a_{16}\xi)f(a_{17}(a_{18}-\xi))
\end{eqnarray}
with
$
\zeta=\log(p_r/\mbox{dyn}\, \mbox{cm}^{-2}), \, \xi=\log(\rho/\mbox{g}\,\mbox{cm}^{-3})\,,
$
where $\rho$ is the neutron matter density and $f(x)=[\exp(x)+1]^{-1}$.
The values of the coefficients $a_{i}$ can be found in Ref.~\cite{Haensel:2004nu}. The corresponding analytical approximations for the FPS EoS and the BSk21 EoS
can be found in Refs.~\cite{Haensel:2004nu} and~\cite{Potekhin:2013qqa}, respectively.

Also, it is necessary to choose the gravitational Lagrangian.
 In this paper we work within the framework of $R^2$ gravity, for which
\begin{equation}
\label{h_R_part_pow}
f=R+\alpha h(R)\equiv R+\alpha R^2.
\end{equation}
The value of the free parameter  $\alpha$ appearing here should be constrained from observations.
In the case of  $R$-squared gravity there are two constraints on
 $\alpha$. First, in the weak-field limit, it is constrained by binary pulsar data as $|\alpha| \lesssim 5\times 10^{15} \text{cm}^2$~\cite{Naf:2010zy}. Second,
 in the strong gravity regime, the constraint is $|\alpha| \lesssim 10^{10} \text{cm}^2$~\cite{Arapoglu:2010rz}. Here, we follow Ref.~\cite{Astashenok:2017dpo}
and take two different values $\alpha = -5 \times 10^{10} \text{cm}^2$ and $\alpha = -20 \times 10^{10} \text{cm}^2$.
(Notice that since here we employ the metric signature distinct from that of Ref.~\cite{Astashenok:2017dpo},  we take opposite signs for $\alpha$
as compared with those used in~\cite{Astashenok:2017dpo}.) If one takes another sign of $\alpha$,
it can lead to the appearance of  ghost modes and
instabilities in the cosmological context~\cite{Barrow:1983rx}
and result in the oscillating behavior of $R$ outside the star,
which appears to be unacceptable in constructing realistic models of neutron stars (for a detailed discussion, see Ref.~\cite{Astashenok:2017dpo}).

\begin{figure}[b]
\centering
  \includegraphics[height=12cm]{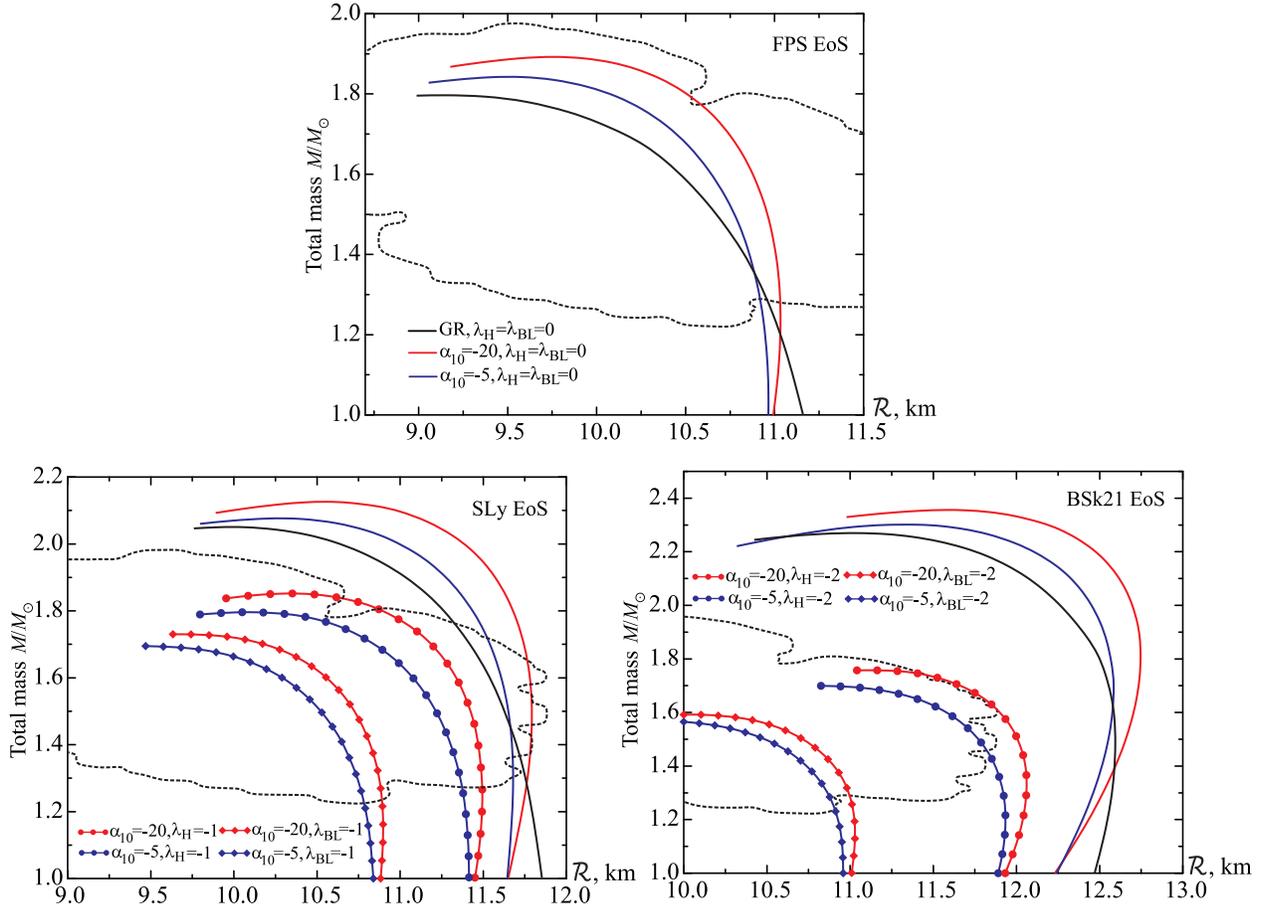}
\caption{The mass-radius relations for the neutron stars in GR and in the MGT for different values of $\alpha_{10}$
(the subscript 10 by $\alpha$ denotes that it is given in units  $10^{10} \text{cm}^2$) and of the anisotropy
parameters $\lambda_{\text{H}}$ and $\lambda_{\text{BL}}$.
The dashed contours depict the region of the observational constraints~\cite{Ozel:2010fw}.
The labeling of the curves for the isotropic configurations given in the top panel refers to the bottom panels as well.
}
\label{fig_M_R}
\end{figure}

\begin{figure}[t]
\centering
  \includegraphics[height=15cm]{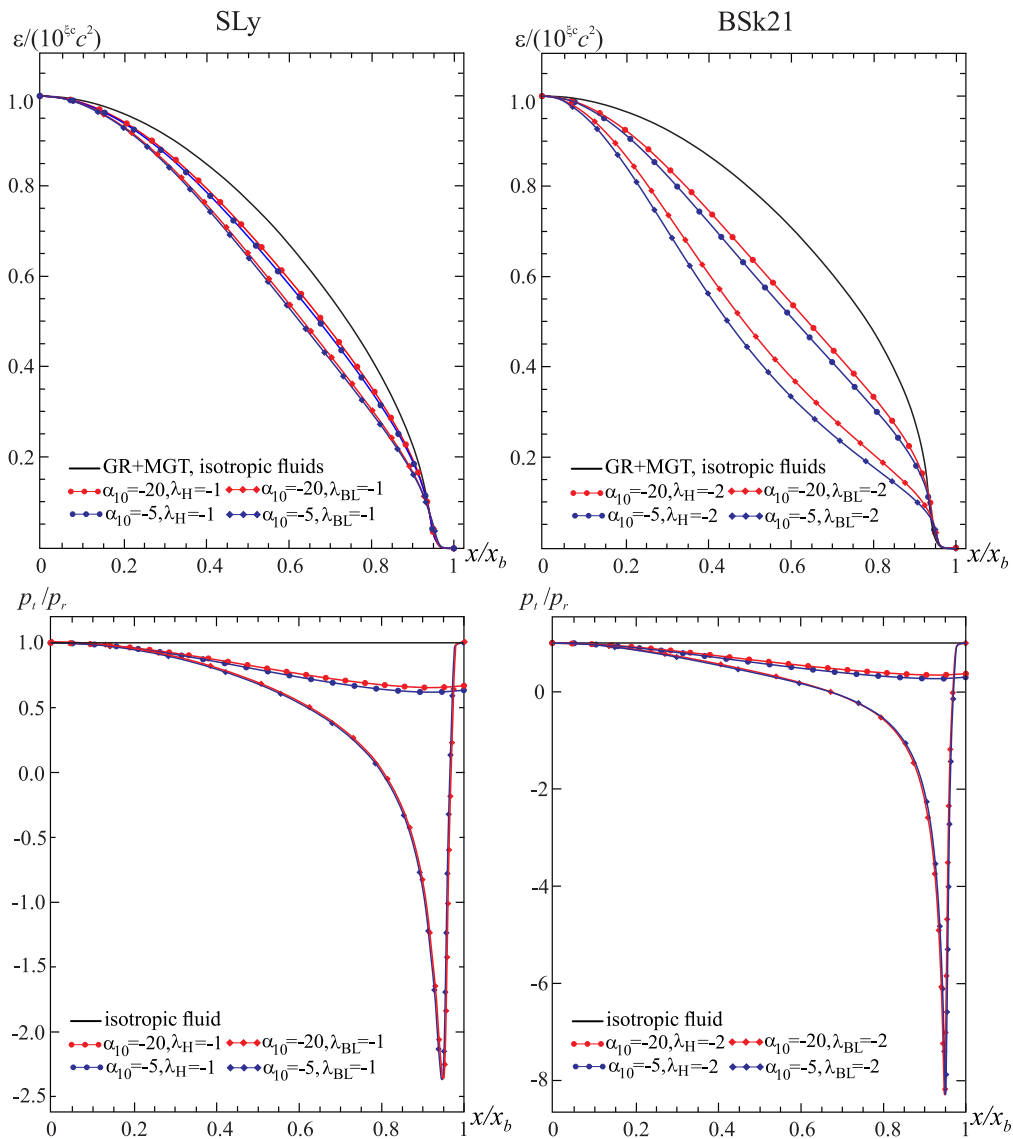}
\caption{The radial distributions of the energy density $\varepsilon$ (expressed in units of the central energy density) and of the ratio $p_t/p_r$ of the tangential pressure to the radial one
for the configurations with the mass $M\approx 1.55 M_\odot$. The left graphs are plotted for the
 SLy EoS,  and the right graphs -- for the BSk21 EoS. The radial coordinate $x$ is normalized to the radius of the fluid $x_b$.
 In the isotropic case, the curves for the energy density of the systems from GR and the MGT nearly coincide;
 they are shown by one solid black line.
}
\label{fig_Energy_pressures}
\end{figure}

For numerical calculations,
it is convenient to rewrite Eqs.~\eqref{scal_cur_eq}, \eqref{mod_Einstein-11_gen}, \eqref{mass_eq}, and \eqref{conserv_gen} in terms of dimensionless variables
\begin{equation}
\label{dimless_var_NS_mod}
x=\frac{r}{L}, \quad \Sigma=R L^2, \quad v(x)=\frac{M(r)}{4\pi 10^{\xi_c} L^3},
\end{equation}
where $L$ is some characteristic length (which is taken to be $L=10^6\,\text{cm}$ in the numerical calculations presented below)
and $\xi_c$ is the central density. Using these variables,
one can get the following set of dimensionless equations for $h(R)$ in the form of \eqref{h_R_part_pow} and the anisotropy factor \eqref{anis_fact_1}:
\begin{eqnarray}
\label{scal_cur_part}
&& \Sigma^{\prime\prime}=- \left[\frac{2}{x}+\frac{1}{2}\left(\nu^\prime-\frac{v^\prime-v/x}{\delta\, x(1-\mu)}\right)\right] \Sigma^\prime
-\frac{1}{6\bar{\alpha}\left(1-\mu\right)}\Big\{b\left[c^2 10^\xi-\left(3+2\lambda_{\text{H}}\mu\right)10^\zeta\right]+\Sigma\Big\},
\\
\label{mod_Einstein-11_part}
&&-(1-\mu)\left(\frac{1}{x^2}+\frac{\nu^\prime}{x}\right)+\frac{1}{x^2} \nonumber
\\
&&+\bar{\alpha}
\Big\{
2 \Sigma\left[
\frac{1}{2}\Sigma-(1-\mu)\left(\frac{1}{x^2}+\frac{\nu^\prime}{x}\right)+\frac{1}{x^2}
\right]-\frac{1}{2}\Sigma^2
-2 (1-\mu) \left(\frac{1}{2}\nu^\prime+\frac{2}{x}\right)\Sigma^\prime
\Big\}=-b 10^\zeta,
 \end{eqnarray}
 \begin{eqnarray}
\label{mass_eq_part}
&&v^\prime=x^2 10^{\xi-\xi_c}- \bar{\alpha} x^2
\Big\{
2\times 10^{\xi-\xi_c}\Sigma+\frac{\delta}{2}  \Sigma^2
-\delta(1-\mu)\left[2\Sigma^{\prime\prime}-\left(\frac{v^\prime-v/x}{\delta\, x-v}-\frac{4}{x}\right)\Sigma^\prime
\right]
\Big\},
\\
\label{conserv_part}
&&\xi^\prime=\frac{1}{ \ln{10}}\frac{1}{d\zeta/d\xi}\left[
-\frac{1}{2}\left(c^2 10^{\xi-\zeta}+1\right)\nu^\prime+\frac{2}{x}\lambda_{\text{H}}\mu
\right],
\end{eqnarray}
where the prime denotes now differentiation with respect to $x$, $\delta=c^2/\left(8\pi G L^2 10^{\xi_c}\right)$,  $\mu=v/(\delta \,  x)$, $\bar{\alpha}=\alpha/L^2$, $b=8\pi G L^2/c^4$.
In a similar way one can derive dimensionless equations for the anisotropy factor \eqref{anis_fact_2}
(we do not show them here to avoid overburdening the text).

These equations are to be solved subject to the boundary conditions given in the neighborhood of the center by the following expansions:
\begin{equation}
\label{bound_mod_Ein}
\xi\approx\xi_c+\frac{1}{2}\xi_2 x^2, \quad \nu\approx\nu_c+\frac{1}{2}\nu_2 x^2, \quad v\approx \frac{1}{3} v_3 x^3, \quad
\Sigma\approx\Sigma_c+\frac{1}{2}\Sigma_2 x^2,
\end{equation}
where the expansion coefficients $\xi_2, \nu_2, v_3,$ and $\Sigma_2$
are determined from Eqs.~\eqref{scal_cur_part}-\eqref{conserv_part}. The central value of the scalar curvature $\Sigma_c$
is chosen so that asymptotically $\Sigma(x\to \infty)\to 0$. In turn, the integration constant $\nu_c$ is fixed by
requiring that the spacetime be asymptotically flat, i.e., $ e^\nu= 1$ at infinity.

Using these boundary conditions, we numerically integrate Eqs.~\eqref{scal_cur_part}-\eqref{conserv_part}.
The integration is performed  from the center (i.e., from $x\approx 0$) to the point $x_b$,
where the neutron matter density decreases to the value
 $\rho_b \approx 10^6 \text{g cm}^{-3}$.
 We take this point to be a boundary of the star.
This density corresponds to the outer boundary of a neutron star crust
up to which the EoSs used here remain valid~\cite{Haensel:2004nu,Potekhin:2013qqa}.
In turn, at $x>x_b$ the neutron matter is absent, i.e., $\rho=p_r=p_t=0$. In GR,
this would correspond to the fact that the scalar curvature $\Sigma=0$.
But this is not the case in the MGT considered here: there exists an external
 gravitational sphere around the star in which $\Sigma\neq 0$.
Consistent with this, the internal solutions must be matched with the external
ones at the boundary of the fluid. This is done by equating the
corresponding values of both the metric functions and the scalar curvature.

For negative $\alpha$'s used in the present paper,
the scalar curvature is  damped exponentially fast outside the star as
$
\Sigma\sim \exp{\left(-x/\sqrt{6|\bar \alpha|}\right)}/ x.
$
This enables one to introduce a  well-defined notion for the gravitational (ADM) mass through Eq.~\eqref{metr_g11},
unlike the case of  positive $\alpha$'s, where $\Sigma$ demonstrates an oscillating behavior~\cite{Astashenok:2017dpo}.

The results of numerical calculations are shown in Figs.~\ref{fig_M_R} and~\ref{fig_Energy_pressures}.
Figure~\ref{fig_M_R} shows the $M-{\cal R}$ relations for typical values of the star's mass $M\sim (1-2)\, M_\odot$.
The dashed contours correspond to the region of observational
constraints obtained for three neutron stars~\cite{Ozel:2010fw}.
It is seen from Fig.~\ref{fig_M_R} that in the case of the isotropic fluid
 (when $\lambda_{\text{H}}= \lambda_{\text{BL}}=0$), the behavior of the $M-{\cal R}$ curves is as follows:
 for the soft FPS EoS a considerable part of the curves lies in the region of the observational
constraints both in GR and in the MGT. For the stiffer  SLy EoS, only an insignificant part of the curves
lies within the observational constraints.
Last, in the case of the stiffest
 BSk21 EoS, the corresponding curves at $\lambda_{\text{H}}= \lambda_{\text{BL}}=0$ do not fall into the observational
constraints at all. Thus, within the assumption of isotropy of the neutron fluid, such a stiff EoS cannot be used to model objects satisfying the current observations
on the $M-{\cal R}$ relations.

\begin{table}
 \caption{Characteristics of the configurations under investigation.
 The radius of surface of the fluid
 (the radius of the star) ${\cal R}$
and the radius of the innermost stable
 circular orbit $r_{\text{ISCO}}$ (see Sec.~\ref{sec_thin_acc_disk})
are given in kilometers. $\rho_c$ is the central density of the neutron matter.
The total mass of all systems  $M\approx 1.55 M_\odot$.
}
\vspace{.3cm}
\begin{tabular}{p{1.5cm}p{1.5cm}p{1.5cm}p{3cm}p{1.5cm}p{1.5cm}}
\hline \\[-5pt]
$\alpha_{10}$ &  $\lambda_{\text{H}}$  & $\lambda_{\text{BL}}$ & $ \rho_c \times 10^{15}, $ $\text{g cm}^{-3}$  &$ {\cal R},$ km  &
 $ r_{\text{ISCO}},$ km  \\[2pt]
\hline \\[-7pt]
\multicolumn{6}{c}{FPS EoS} \\[2pt]
\end{tabular}
\begin{tabular}{p{1.5cm}p{1.5cm}p{1.5cm}p{3cm}p{1.5cm}p{1.5cm}}
\hline \\[-15pt]
0&	0&	0&	1.57&	10.60&	13.72\\
-5&	0&	0&	1.55&	10.76&	14.35\\
-20&	0&	0&	1.51&	10.96&	13.74\\
\hline
\multicolumn{6}{c}{$\phantom{\Big(}$ SLy EoS}\\[2pt]
\hline \\[-15pt]
0&	0&	0&	1.11&	11.60&	13.72\\
-5&	0&	0&	1.12&	11.67&	14.48\\
-20&	0&	0&	1.12&	11.83&	13.77\\
-5&	-1&	0&	1.48&	11.18&	14.46\\
-20&	-1&	0&	1.41&	11.41&	13.82\\
-5&	0&	-1&	2.00&	10.45&	14.40\\
-20&	0&	-1&	1.88&	10.69&	13.77\\

\hline
\multicolumn{6}{c}{$\phantom{\Big(}$ BSk21 EoS}\\[2pt]
\hline\\[-15pt]
0&	0&	0&	0.80&	12.60&	13.72\\
-5&	0&	0&	0.82&	12.61&	14.71\\
-20&	0&	0&	0.82&	12.73&	13.77\\
-5&	-2&	0&	1.42&	11.71&	14.57\\
-20&	-2&	0&	1.29&	11.99&	13.79\\
-5&	0&	-2&	3.16&	10.21&	14.32\\
-20&	0&	-2&	2.63&	10.57&	13.78\\
\hline
\end{tabular}
\label{tab1}
\end{table}

Hence we see that, as already pointed out in Ref.~\cite{Ozel:2010fw},
in the case of modeling NSs within the framework of GR the observational data imply that matter supporting the NSs should be described by one of soft EoSs
(for example, the FPS EoS used here or the AP4 EoS considered in~\cite{Ozel:2010fw}). Our purpose here is to try to modify the system
in such a way that, keeping in mind that the presence of the anisotropic pressure is possible in principle,
the $M-{\cal R}$ curves would also fall  into the region of observational
constraints when one uses more stiff EoSs. As the calculations indicate, this can be done only
for negative values of
 $\lambda_{\text{H}}, \lambda_{\text{BL}}$, which means that the tangential pressure $p_t$ is less than the radial pressure
  $p_r$ [see the expressions~\eqref{anis_fact_1} and ~\eqref{anis_fact_2}].
 Figure~\ref{fig_M_R} shows the corresponding $M-{\cal R}$  relations for two values of
  $\lambda_{\text{H}}, \lambda_{\text{BL}}=-1$ or $-2$, which allow us to get configurations with characteristics that are
  more or less acceptable from the observational point of view.

Apart from the changes in the $M-{\cal R}$ relations, the presence of the anisotropy also leads to changes in
the distributions of the energy density and pressures along the radius of the configuration.
These changes are illustrated in Fig.~\ref{fig_Energy_pressures}
for the configurations with the fixed mass  $M\approx 1.55 M_\odot$.
This choice of the mass is made, first, because
it can be realized for all EoSs and values of the parameters
$\alpha, \lambda_{\text{H}}, \lambda_{\text{BL}}$ used in the paper, and second,
because the configurations with such a mass lie in or close to the observationally allowed region
(except the isotropic systems supported by the BSk21 EoS). As one can see from Fig.~\ref{fig_Energy_pressures},
the profiles of the energy density distributions for the isotropic configurations practically coincide
for all EoSs in question (including the FPS EoS, which is not shown in Fig.~\ref{fig_Energy_pressures}),
regardless of whether the modeling is carried out within the framework of GR or in the MGT.
In the presence of the anisotropy, the matter concentrates toward the center
when at the given relative radius $x/x_b$ the energy density is smaller than that in the isotropic case.
In turn, the radius of the anisotropic configurations decreases as compared with that of the isotropic systems (see Table~\ref{tab1}).
All this is a consequence of the fact that in the presence of the anisotropy
greater central densities of the matter must be taken to get the required fixed mass.

As for the ratios of the pressures (see Fig.~\ref{fig_Energy_pressures}),
their distributions along the radius are basically determined by the actual type of the anisotropy and by the form of the EoS,
and not by the theory of gravity that is used to model the star
(a weak dependence on the value of $\alpha$).
Moreover, if in the case of using the anisotropy factor \eqref{anis_fact_1} the difference between
$p_t$ and $p_r$ changes relatively slowly along the radius, in the case of the anisotropy factor from \eqref{anis_fact_2},
the ratio $p_t/p_r$ changes considerably more rapidly,
especially  in the external regions of the star.
It is also interesting to note that the tangential pressure in the external regions becomes even negative; i.e., it plays the role of tension, similar to that appearing in solid bodies at their stretching.

\section{Thin accretion disk}
\label{sec_thin_acc_disk}

In this section we consider the process of accretion of test particles onto our configurations.
The purpose is to clarify the differences between the neutron stars constructed within the framework of the MGT and those from GR
as regards the observational manifestations associated with the accretion process.

\subsection{Description of the model}

We will closely follow the work of Page and Thorne~\cite{Page:1974he}, who studied the relativistic model
of thin-disk accretion onto a black hole. In doing so, we will not consider the process of the infall of accreting matter onto the surface
of the NSs and changes in the emission spectra associated with such a process, but consider only
phenomena related to the accretion disk. During the accretion process,
a fraction of the heat converts into electromagnetic radiation and cools down the disk.
The analysis of the resulting spectrum of emission enables one to reveal the distinguishing features of configurations onto which the accretion takes place.

Within the framework of the model of Ref.~\cite{Page:1974he},
the following characteristics are assumed.
(i) The accretion disk has a negligible influence on an external spacetime geometry
(black hole geometry in~\cite{Page:1974he}). (ii) The disk resides in the equatorial plane of the central object.
(iii) The disk is thin; i.e., its thickness is much smaller than its radius. (iv)~The physical quantities describing the gas
in the disk are averaged over a characteristic time interval $\Delta t$ and the azimuthal angle $\Delta \varphi=2\pi$.
(v) Within the disk, there is a heat flow only in the vertical direction.

Using these assumptions and the laws of conservation of rest mass, angular momentum, and energy,
one can obtain the following formula for the time-averaged flux of radiant energy flowing out of the upper or lower side of the disk~\cite{Page:1974he}:
\begin{equation}
\label{flux}
F(r)=-\frac{\dot{M}_0 c^2}{4\pi \sqrt{-g}}\frac{\Omega_{,r}}{\left(\bar{E}-\Omega\bar{L}\right)^2}\int_{r_{\text{ISCO}}}^{r}
\left(\bar{E}-\Omega\bar{L}\right)\bar{L}_{,r} dr.
\end{equation}
Here $\bar{L}$, $\bar{E}$, and $\Omega$ are  the specific angular momentum,  the
specific energy, and the angular velocity of particles moving in circular orbits around the central body, respectively;
$\dot{M}_0=\text{const.}$ is the time-averaged rate at which rest mass flows inward through the disk.
The subscript $,r$ denotes the derivative with respect to $r$.
The lower limit of integration $r_{\text{ISCO}}$ is chosen to be the innermost stable
 circular orbit (ISCO) from which
the accreting matter falls freely onto the central object.

All quantities appearing in Eq.~\eqref{flux} depend on the
radial coordinate $r$ only. According to the above assumptions (ii) and (iii),
in order to describe the accretion process, one can
employ the following cylindrical metric in the neighborhood of  the equatorial plane ($|\theta-\pi/2|\ll 1$):
\begin{equation}
\label{metr_equator}
ds^2=e^{2\gamma}(d x^0)^2-e^{2\alpha}dr^2-e^{2\beta}d\varphi^2-dZ^2,
\end{equation}
where the functions $\alpha, \beta, \gamma$ depend on $r$ only.
[This metric can be obtained from the general spherically symmetric line element by replacing the usual angular coordinate $\theta$
by $Z=e^\beta\cos{\theta}\approx e^\beta(\theta-\pi/2)$.]

Using this metric, we can integrate the geodesic equation. In considering
timelike geodesics for massive particles, one can derive the following formulas for
the specific energy and the specific angular momentum: $\bar{E}=c^2 e^{2\gamma} \dot{t}$ and
$\bar{L}=e^{2\beta}\dot\varphi$, where the dot denotes  the derivative with respect to the proper time $\tau$
along the path.

Next, substituting the above expressions for  $\bar{E}$ and  $\bar{L}$ into
 a first integral of the geodesics equations $g_{\mu\nu}\dot{x}^\mu\dot{x}^\nu=c^2$,
  one can derive the following ``energy'' equation for a particle
\begin{equation}
\label{energ_cons_part}
\frac{\bar{E}^2}{c^2}=e^{2(\alpha+\gamma)}\dot{r}^2+V_{\text{eff}}
\end{equation}
with the effective potential
\begin{equation}
\label{eff_poten}
V_{\text{eff}}(r)=e^{2\gamma}\left(c^2+e^{-2\beta} \bar{L}^2\right).
\end{equation}

When one considers a circular motion in the equatorial plane, it is obvious that $r=\text{const.}$ Correspondingly, Eq.~\eqref{energ_cons_part} yields
$V_{\text{eff},r}=0$. Using this together with Eq.~\eqref{energ_cons_part} and taking into account the definition of the angular velocity
$\Omega=d\varphi/dt$, one can get the following expressions:
\begin{eqnarray}
\label{ang_vel}
&&\Omega=c e^{\gamma-\beta}\sqrt{\frac{\gamma_{,r}}{\beta_{,r}}},\\
\label{ang_mom}
&&\bar{L}=\frac{c \Omega e^{2\beta}}{\sqrt{c^2 e^{2\gamma}-e^{2\beta}\Omega^2}},\\
\label{energ_specif}
&&\bar{E}=\frac{c^3 e^{2\gamma}}{\sqrt{c^2 e^{2\gamma}-e^{2\beta}\Omega^2}}.
\end{eqnarray}
Substituting them into \eqref{flux}, one can derive a radial distribution of the energy flux.

Let us now rewrite the obtained formulas in terms of the dimensionless variables used above.
The characteristic size of the systems under consideration is $L$ from \eqref{dimless_var_NS_mod}.
According to Eq.~\eqref{metric_schw}, the metric functions entering
\eqref{metr_equator} are
 $\gamma=\nu/2, \alpha=\lambda/2$, and $e^\beta=r$. Then  Eqs.~\eqref{ang_vel}-\eqref{energ_specif} yield
 \begin{equation}
\label{func_for_mixed}
\Omega=\frac{c\, e^{\nu/2}}{L} \sqrt{\frac{\nu'}{2 x}},\quad
\bar{L}=c \,L\sqrt{\frac{x^3 \nu'}{2-x \nu'}}, \quad
\bar{E}=c^2\sqrt{\frac{2 e^\nu  }{2-x \nu'}}.
\end{equation}
Here the prime denotes differentiation with respect to $x$ from  \eqref{dimless_var_NS_mod}.
Substituting these expressions into Eq.~\eqref{flux}, one can find the flux for the systems under consideration:
 \begin{equation}
\label{flux_mixed}
F(x)=-\frac{\dot{M}_0 c^2}{4\pi L^2}\frac{\Omega^\prime}{e^{(\nu+\lambda)/2} x \left(\bar{E}-
\Omega\bar{L}\right)^2}\int_{x_{\text{ISCO}}}^{x}
\left(\bar{E}-\Omega\bar{L}\right)\bar{L}^\prime dx.
\end{equation}
Note that  $\Omega,\bar{L}$, and $\bar{E}$ appearing in Eq.~\eqref{flux_mixed}
are taken from \eqref{func_for_mixed} without the dimensional coefficients $c$ and $L$.

In turn, the effective potential \eqref{eff_poten} takes the form
\begin{equation}
\label{eff_poten_mixed}
V_{\text{eff}}(x)=c^2\frac{2 e^\nu  }{2-x \nu'}.
\end{equation}

Using this, the circular orbits are obtained from the condition $d V_{\text{eff}}/dx=0$,
and the orbits are stable or unstable if
 $d^2 V_{\text{eff}}/dx^2>0$ or $d^2 V_{\text{eff}}/dx^2<0$, respectively.

Consider now the question of the spectrum emitted from the surface of the disk. For this purpose,
we have to determine the spectrum
emitted locally at each point of the disk and then carry out the integration over the whole disk surface.
To do this, we start from the assumption that the disk is optically thick; i.e., it is assumed that each element of the disk radiates as a black
body with temperature $T(r)$. Then, using  the above flux, one can find this temperature via the formula
$F(r)=\sigma_{\text{SB}} T^4(r)$,
where $\sigma_{\text{SB}}$ is the Stefan-Boltzmann constant.
Using this temperature distribution, one can calculate
the total energy radiated from both sides of the disk
 at frequency  $\omega$ as
$$
S(\omega)=2 \int I(\omega) d S_d
\quad
\text{with}  \quad I(\omega)=\frac{\hbar \omega^3}{2\pi^2 c^2}\frac{1}{e^{\hbar\omega/k_B T}-1},$$
where $I(\omega)$ is the Planck distribution function, and
$k_B$ is the Boltzmann constant. The surface area of the disk
$S_d$ appearing in the above formula is $$S_d=2\pi\int_{r_{in}}^{r_{out}}e^{\beta} dr,$$
where $r_{in}$ and $r_{out}$ are the inner and outer radii of the disk
 [recall that here $\beta$ is the metric function from \eqref{metr_equator}].

Using the obtained expressions and the dimensionless variables \eqref{dimless_var_NS_mod}, one can find
\begin{equation}
\label{energ_disk_gen}
S(\omega)=\frac{2\hbar}{\pi c^2}\omega^3 \int_{r_{in}}^{r_{out}}\frac{e^{\beta}}{e^{\hbar\omega/k_B T}-1}dr=
\frac{2\hbar}{\pi c^2}L^2\omega^3 \int_{x_{in}}^{x_{out}}\frac{x}{e^{\hbar\omega/k_B T}-1}dx,
\end{equation}
where we have taken into account that $e^\beta=r$.
If the accretion disk is inclined with respect to an observer at angle $i$ (i.e., the angle between the line of sight and the normal to the disk),
the measured energy is calculated by multiplying the above expression  by $\cos{i}$.
We emphasise that the formula
\eqref{energ_disk_gen} gives the amount of the total energy emitted at the given frequency from the whole disk surface,
but not the distribution of the energy along the radius. This assumes that a distant observer registers this energy at the given frequency.

\subsection{Results of calculations}

Bearing in mind that our aim is to reveal the observational differences between the NSs constructed within the framework of GR and those from the MGT,
we perform here a comparison of the systems with the same masses. As in Sec.~\ref{num_calc},
we consider the configurations with the mass $M\approx 1.55 M_\odot$.

Note that since NSs have a material surface,
then as accreting matter falls onto such a surface, it will
emit a luminosity of the same order as that emitted by the disk~\cite{Novikov:1973}.
If the total luminosity becomes of the order of the ``Eddington limit'',
$L_{\text{Edd}}\sim \left(10^{38} \text{erg/sec}\right) \left(M/M_\odot\right)$,
then radiation pressure will destroy the disk.
In this case the standard thin disk model by Shakura and Sunyaev \cite{Shakura:1972te} employed here is not already applicable.
This assumes that the accretion rate should be very sub-Eddington
 (i.e., the total luminosity should be much less than $L_{\text{Edd}}$). For this case the accretion rate
$$
\dot{M}_0 \ll \dot{M}_{\text{Edd}}\sim \left(10^{-8} M_{\odot}/\text{yr}\right) \left(M/M_\odot\right).
$$

The results of calculations presented below are obtained for the mass accretion rate $\dot{M}_0 =10^{-12} M_{\odot}/\text{yr}$ \cite{Shakura:1972te}.
The outer radius of the accretion disk is taken to be $r_{out}=10^3 G M/c^2$~\cite{Staykov:2016dzn}.  The inner edge of the disk is on the ISCO,
i.e., $r_{in}=r_{\text{ISCO}}$, whose numerical values for the systems under consideration are given in Table~\ref{tab1}.

\begin{figure}[b]
\centering
  \includegraphics[height=6cm]{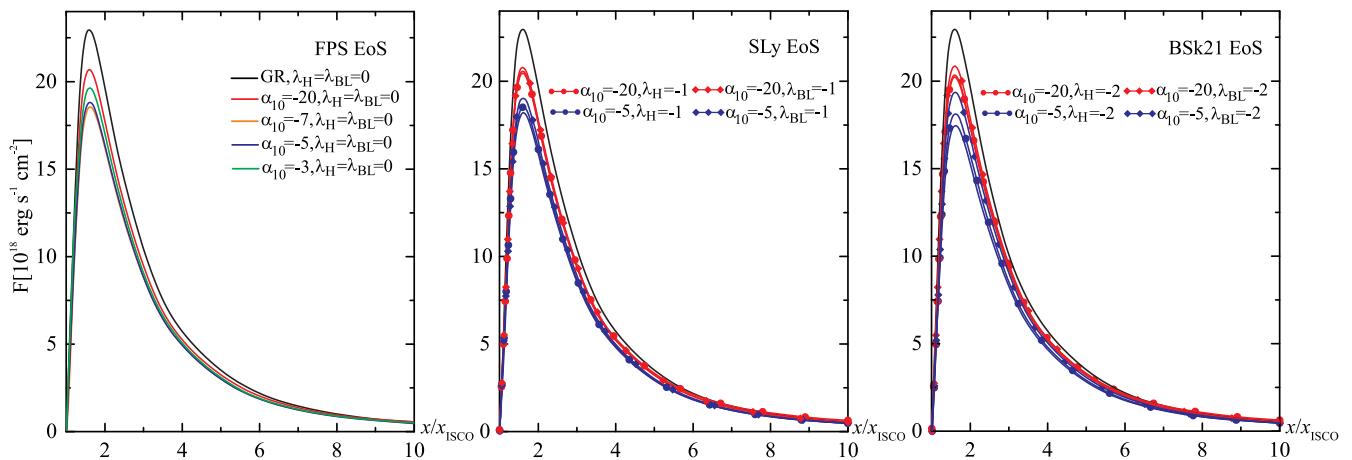}
\caption{The fluxes are shown as functions of the relative radius $x/x_{\text{ISCO}}$.
The labeling of the curves for the isotropic configurations given in the left panel refers to the other panels as well.
}
\label{fig_fluxes}
\end{figure}

The corresponding graphs for the electromagnetic flux
are plotted in Fig.~\ref{fig_fluxes}.
For purposes of comparison, it appears more convenient to work in relative units where
the radial coordinate $x$ is  normalized to $x_{\text{ISCO}}$. Then one can see from
Fig.~\ref{fig_fluxes} that the flux reaches its maximum magnitude always near the inner edge of the accretion disk.
Comparing the GR and MGT results, we see that in the MGT the fluxes are always smaller, regardless of the EoS used here, as well as the magnitude and the form of the anisotropy.
The maximum difference $\sim 25\%$ is reached in the case of the isotropic fluid described by the stiff BSk21 EoS (for the MGT with $\alpha_{10}=-5$).

Notice also the following properties of the systems under consideration:
\begin{itemize}
  \item Within the framework of GR, the distributions of the flux along the radius are practically independent of the EoSs  used here. At the same time, in the MGT, the softer EoS ensures the greater fluxes.
  \item The maxima of the fluxes are always located at approximately the same relative radius $x/x_{\text{ISCO}}$.
  \item As the parameter $\alpha$ increases (in modulus), the flux at first decreases and then starts to increase.
  We have demonstrated this in Fig.~\ref{fig_fluxes} for the case of the FPS EoS by adding two extra graphs for
   $\alpha_{10}=-3$ and $\alpha_{10}=-7$. Similar behavior of the flux also takes place for the other EoSs used here.
 When the parameter $\alpha$ increases (in modulus) further, the flux becomes even larger than that in GR
 (in this connection, see Ref.~\cite{Staykov:2016dzn} where the case of extremely large values of $\alpha_{10}\approx -2\times 10^4$ has been considered).
However, as pointed out earlier, we do not consider here such large $\alpha$'s, remaining within the constraints imposed by the strong gravity regime~\cite{Arapoglu:2010rz}.
  \item The presence of the fluid anisotropy results in the increase of the fluxes as compared with the isotropic case.
  \end{itemize}

\begin{figure}[t]
\centering
  \includegraphics[height=6.5cm]{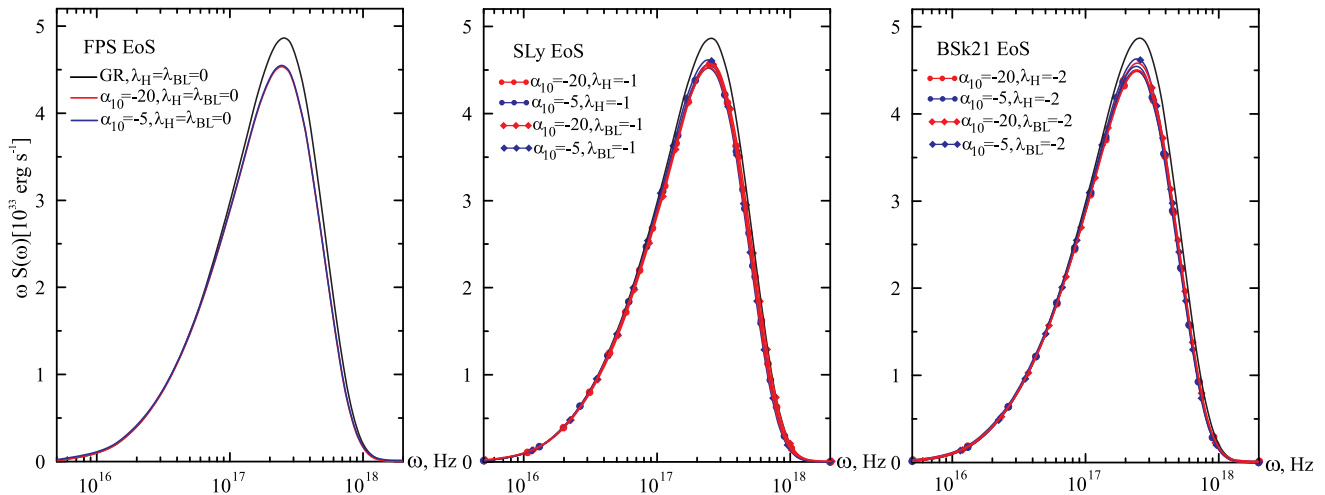}
\caption{The total energy emitted from the disk for a mass accretion rate
  $\dot{M}_0=10^{-12}M_{\odot}/\text{yr}$ and the outer radius of the accretion disk $r_{out}=2.65 \times 10^3 \text{km}$ corresponding to $M=1.55 M_\odot$.
The labeling of the curves for the isotropic configurations given in the left panel refers to the other panels as well.
}
\label{fig_energ_rad_disk}
\end{figure}

Next, the results of calculations of the emission spectrum  from the
formula \eqref{energ_disk_gen} for the X-ray band are shown in Fig.~\ref{fig_energ_rad_disk}.
One can see that the systems in GR and in the MGT have maxima in the emission spectrum
in approximately the
same frequency band. At the same time, as in the case of the flux,
 the radiated energy of the systems in the MGT is less than that of the configurations from GR,
 and the maximum difference reaches the order of $8\%$.

Finally, one can estimate the efficiency of energy radiation, $\epsilon$, in an accretion disc.
The maximum efficiency is of the order of the ``gravitational binding energy''
at the ISCO (i.e., the energy which is lost by a particle when it moves
from infinity to the lowest orbit) divided by the rest mass energy of the particle.
Then, using the expressions for the specific energy $\bar E$ from \eqref{func_for_mixed}, the efficiency is $\epsilon=1-\bar{E}(x_{\text{ISCO}})/c^2$.
Using this expression, we have found that for all configurations under consideration the efficiency of the conversion of the accreted mass into radiation
lies within the range  $5.3\%\lesssim \epsilon \lesssim 5.7\%$. These magnitudes of $\epsilon$ are close to those typical for static (nonrotating) black holes and neutron stars in GR.

\section{Conclusion}
\label{concl}

Neutron stars are objects whose structure and physical characteristics are largely determined by their own strong gravitational fields.
Possessing a number of unique properties,  such stars are
characterized by a sufficiently large variety of observational manifestations,
and one is able to use them to test the correctness of various theoretical models of
extreme states of matter. And conversely, the development of theoretical models of matter at high densities and pressures is a necessary step in constructing
models of NSs that agree sufficiently with observational data.

In the present paper we have studied static NSs within the framework of $R^2$ gravity.
Our purpose was to construct objects whose characteristics would be consistent with the current observational data
on the neutron star mass-radius relation.
The modeling has been carried out using the well-known realistic EoSs describing neutron matter at high densities.
For the isotropic configurations, we showed that both in GR and
in the MGT  the  $M-{\cal R}$ curves agree with the observations only if one uses a soft EoS (the FPS EoS in our case).
 If one intends to employ more stiff EoSs (the SLy and BSk21 EoSs in our case),
 these curves already go beyond the observational constraints,
 and in the MGT these deviations are even stronger than those in GR.

To address this problem,
we have introduced the anisotropy of neutron matter given in two different forms, \eqref{anis_fact_1} and \eqref{anis_fact_2},
which take into account both the local properties of the matter (through pressure) and  the  quasilocal properties of the configuration (through compactness).
By choosing particular values of the anisotropy parameters, we showed that it is possible to shift
the  $M-{\cal R}$ curves to the region of the observational constraints (see Fig.~\ref{fig_M_R}).
We thus demonstrated the possibility in principle of constructing realistic models of NSs using any (whether soft or stiff) EoSs.

Aside from the influence on the $M-{\cal R}$ relation, the presence of the anisotropy leads to considerable changes in the radial distributions of the energy density and pressure of the neutron matter
(see Fig.~\ref{fig_Energy_pressures}). In particular, the greater (in modulus) the magnitudes of the anisotropy parameters,
the greater the concentration of the matter toward the center. At the same time, the difference between the tangential and radial pressures is
basically determined by the actual type of the anisotropy and by the form of the EoS, and not by the theory of gravity within which the modeling is carried out.
Moreover, when one takes the anisotropy factor in the form \eqref{anis_fact_2},  the tangential pressure becomes negative in the external regions of the star.

Neutron stars constructed within the framework of  the MGT may also possess other marked distinctions as compared with NSs from GR. In particular,
since the external spacetime geometry of NSs in GR differs from that obtained in the MGT,
the motion of test particles will in general be different.
This manifests itself, for example, when one considers the process of accretion of matter onto such configurations.
Then, depending on the particular type of the theory of gravity, both the structure of accretion disks
and their radiant emittance (spectrum) will change.

Consistent with this,  we have considered the process of accretion of test particles onto the NSs with the same masses described in GR and in the MGT.
For this purpose, we have employed the well-known thin accretion disk model of Ref.~\cite{Page:1974he}
within which it was shown that (see Figs.~\ref{fig_fluxes} and~\ref{fig_energ_rad_disk})
\begin{itemize}
  \item As compared with GR, in the MGT, the electromagnetic fluxes radiated from the surface of the accretion disk are always smaller,
  regardless of the EoS used here, as well as the magnitude and the form of the anisotropy (the maximum difference in the flux reaches $\sim 25\%$).
 \item The maxima of the fluxes are always reached near the inner  edge of the accretion disk
 and located at approximately the same relative radius $x/x_{\text{ISCO}}$ both in GR and in the MGT.
   \item Within the framework of GR, the radial distributions of the flux
    are practically independent of the EoSs used here. At the same time, in the MGT, the softer EoS ensures greater fluxes.
   \item As the parameter $\alpha$ increases (in modulus), the flux at first decreases and then starts to increase.
   \item The presence of the fluid anisotropy results in the increase of the fluxes as compared with the isotropic case.
  \item  The systems in GR and in the MGT have maxima  in the emission spectrum in approximately the
same frequency band. The radiated energy of the objects in the MGT is less than that of the configurations from GR
 (the maximal difference is of the order of  $8\%$).
\item  The efficiency of the conversion of the accreted mass into radiation lies within the range  $5.3\%\lesssim \epsilon \lesssim 5.7\%$ (it depends on the specific values of the parameters $\alpha, \lambda_{\text{H}}, \lambda_{\text{BL}}$).
\end{itemize}

Summarizing the obtained results, we have demonstrated the influence that
the effects of modified gravity and the fluid anisotropy have on
(i) the mass-radius relations of the neutron stars and their internal structure and
(ii)~the radiant emittance of the accretion disk.
We have shown that the introduction of the anisotropy enables one to obtain a better agreement of theoretical calculations with the observational data on the  $M-{\cal R}$ relation.
This is especially crucial for the moderately stiff and stiff EoSs, for which the theoretical $M-{\cal R}$ curves pertaining to the isotropic configurations
lie outside the observational constraints both in GR and in the MGT.

It is evident that the obtained results are essentially model dependent and are in general determined by a specific type
of $f(R)$ gravity and by a particular form of modeling  the anisotropy in the system.
In particular, instead of $R^2$ gravity used here, one can consider theories with other forms of nonlinear terms.
For instance, these can be cubic or logarithmic terms, employed in Ref.~\cite{Capozziello:2015yza} to obtain
the mass-radius relations for NSs modeled by various realistic EoSs. The results of Ref.~\cite{Capozziello:2015yza} indicate that the qualitative behavior
of the $M-{\cal R}$ curves remains approximately the same as that observed in $R^2$ gravity.
In this connection one may expect that, working within the framework of different $f(R)$ gravities and varying the anisotropy parameters,
it will also be possible to achieve a good agreement of theoretical calculations with the observational data on the  $M-{\cal R}$ relation.

As for the form of the anisotropy,
there is no fully reliable way at present to determine the true nature and magnitudes of the anisotropy
in realistic superdense configurations. Consequently, the introduction of a specific model for the anisotropy is a delicate issue, which usually amounts to using a phenomenological approach. In such a case, it is
necessary to ensure that the magnitudes of free parameters appearing in such phenomenological models would adequately correspond  to field-theoretical
models of anisotropic matter currently considered in the literature.
In particular, in order to ensure the compatibility of the anisotropy factor in the form \eqref{anis_fact_1}
with the model of anisotropy occurring due to pion condensation \cite{Saw1972}, one must take the anisotropy parameter
 $\lambda_{\text H}$ from  \eqref{anis_fact_1} to be of the order of unity \cite{Doneva:2012rd}
 (the typical value used in the present paper).
This allows the possibility of  ensuring certain reliability of the phenomenological model of anisotropy employed here.

In any case,
 if the matter of neutron stars may possess an anisotropic pressure,
one might expect changes both in the structure of the stars and in the mass-radius relation,
regardless of the specific way in which the anisotropy is modeled.
These changes can in principle be verified observationally, and this can help one to exclude some nonviable approaches in modeling the anisotropy.
Also, astrophysical observations of emission spectra from accretion disks can provide
an opportunity to distinguish the external geometry  of neutron stars described in GR from the one obtained in the MGT.

\section*{Acknowledgments}
The author gratefully acknowledges support provided by Grant No.~BR05236494
in Fundamental Research in Natural Sciences
by the Ministry of Education and Science of the Republic of Kazakhstan.
I am also very grateful to V. Dzhunushaliev for fruitful discussions and comments.

\end{document}